\documentclass[fleqn,usenatbib]{mnras}

\usepackage{newtxtext,newtxmath}
\usepackage{lineno}

\usepackage[T1]{fontenc}

\DeclareRobustCommand{\VAN}[3]{#2}
\let\VANthebibliography\thebibliography
\def\thebibliography{\DeclareRobustCommand{\VAN}[3]{##3}\VANthebibliography}

\usepackage{multicol}
\usepackage{graphicx}	
\usepackage{amsmath}	

\title[uGMRT observations of RS Ophiuchi]{Shock-driven synchrotron radio emission from the 2021 outburst of RS Ophiuchi}

\author[Nayana et al.]{
Nayana A.J.,$^{1,2}$\thanks{E-mail: nayana.aj@iiap.res.in}
G.C. Anupama,$^{2}$
Nirupam Roy$^{3}$
Dipankar P. K. Banerjee$^{4}$
Kulinder Pal Singh$^{5}$
Sonith L.S.,$^{2}$ 
\newauthor
and U. S. Kamath$^{2}$
\\
$^{1}$Department of Astronomy, University of California, Berkeley, CA 94720-3411, USA.\\
$^{2}$Indian Institute of Astrophysics, II Block, Koramangala, Bangalore 560034, India.\\
$^{3}$Department of Physics, Indian Institute of Science, Bangalore 560012, India.\\
$^{4}$Astronomy \& Astrophysics Division, Physical Research Laboratory, Ahmedabad 380009, India.\\
$^{5}$ Department of Physical Sciences, IISER Mohali, Knowledge City, Sector 81, Manauli PO, SAS Nagar, Punjab 140306, India.
}

\date{Accepted XXX. Received YYY; in original form ZZZ}

\pubyear{2023}

\begin{document}
\label{firstpage}
\pagerange{\pageref{firstpage}--\pageref{lastpage}}
\maketitle

\begin{abstract}
We present low-frequency radio observations of the Galactic symbiotic recurrent nova RS Ophiuchi during its 2021 outburst. The observations were carried out with the upgraded Giant Metrewave Radio Telescope (uGMRT) spanning a frequency range of 0.15$-$1.4 GHz during 23$-$287 days post the outburst. The average value of the optically thin spectral index is $\alpha \sim$ $-$0.4 ($F_{\nu} \propto \nu^\alpha$), indicating a non-thermal origin of the radio emission at the observed frequencies. The radio light curves are best represented by shock-driven synchrotron emission, initially absorbed by a clumpy ionized circumbinary medium. We estimate the mass-loss rate of the red giant companion star to be $\dot{M} \sim$ 7.5 $\times$ 10$^{-8}$ $M_{\odot}$\,yr$^{-1}$ for an assumed stellar wind velocity of 20 km/s. The 0.15--1.4 GHz radio light curves of the 2021 outburst are systematically brighter than those of the 2006 outburst. Considering similar shock properties between the two outbursts, this is indicative of a relatively higher particle number density in the synchrotron emitting plasma in the current outburst. 
\end{abstract}




\section{Introduction}

RS Ophiuchi is a Galactic symbiotic recurrent nova system composed of a primary white dwarf that accretes matter from its red giant companion. The nova has undergone seven reported outbursts, at somewhat irregular intervals, in the years 1898, 1933, 1958, 1967, 1985, 2006, and 2021 \citep[][and references therein]{schaefer2010}. The 2021 outburst was discovered on 2021 August 08.93 (UT) in the optical bands \footnote{http://ooruri.kusastro.kyoto-u.ac.jp/mailarchive/vsnet-alert/26131}. The nova brightened to a visual magnitude of 4 mag, which is $\sim$ 1000 times brighter than its quiescent magnitude. Subsequently, the nova showed luminous emission over a wide range of frequencies from radio to the Very High Energy (VHE) $\gamma$-rays. 

$\gamma$-ray emission was detected with Fermi-LAT \citep{cheung2022,zheng2022}, H.E.S.S \citep{aharonian2022}, and MAGIC \citep{acciari2022} telescopes in the GeV-TeV energy regime, marking the first detection of VHE emission from a nova. The GeV to TeV emission has been interpreted under either a hadronic \citep{aharonian2022,acciari2022,zheng2022,diesing2023} or a lepto-hadronic \citep{agnibha2023} particle acceleration scenario. The evolution of the hard X-ray emission, arising from shock interaction between the nova ejecta and the red giant wind, was found to be similar to the 2006 outburst \citep{page2022}. On the other hand, the evolution in the soft X-ray, during the super soft state (SSS), was found to be fainter compared to the 2006 outburst \citep{page2022}. High resolution X-ray spectra observed during $t\sim 18-21$ days with \textit{Chandra} and \textit{XMM Newton} \citep{orio2022} indicated the presence of multiple thermal components over a range of temperatures $T = 0.07-3.4$ keV. 
 
\cite{ness2023} analyzed the X-ray grating spectra of both the 2006 and 2021 outbursts and concluded the low SSS emission in 2021 to be due to higher absorption in the line of sight compared to that of 2006. The early optical spectrum of the 2021 outburst showed emission lines of full width at half maximum $\sim$ 2900 km\,s$^{-1}$ \citep{munari2021atel14840}. The ejecta accelerated to velocities $\sim$ 4700 km\,s$^{-1}$ \citep{mikolajewska2021} in the first two days followed by decreasing line velocities indicating deceleration \citep{pandey2022}. The optical photometric and spectroscopic evolution of the 2021 outburst presented by \cite{munari2021,munari2022} indicate it to be very similar to that of previous outbursts. The morpho-kinematic modeling of H$_{\alpha}$ emission lines suggested a bipolar ejecta geometry \citep{pandey2022} similar to that seen during the 2006 outburst \citep{rupen2008}.

Radio emission was detected from RS Ophiuchi from $\sim$ 1 day post-outburst at multiple frequencies (0.15 $-$ 35 GHz) \citep{williams2021,sokolovsky2021,Nayana2021atel,deruiter2023}. 
\cite{munari2022vlbi} presented the European Very Large Baseline Interferometry Network (EVN) observations of RS Oph at 5 GHz and reported the total extension of the radio-emitting region to be $\sim 90$ mas at $t \sim 34$ days. Radio emission from novae could be thermal free-free emission from hot ionized ejecta and/or non-thermal synchrotron emission from shocks. Low-frequency radio emission are synchrotron-dominated and are critical in understanding the properties of circumbinary medium (CBM) and shocks. 

In this paper, we present extensive low-frequency (0.15 $-$ 1.4 GHz) radio observations and modelling of RS Ophiuchi during its 2021 outburst. The paper is organized as follows: We present the radio observations and data reduction in \S \ref{sec:obs}. Radio light curves and spectra are presented in \S \ref{sec:radio-lcs-spectra} and the radio emission model is explained in \S \ref{sec:radio-model}. We discuss our results in \S \ref{sec:comparison-1985,2006 and 2021} and \S \ref{sec:comprehensive}. We summarize our findings in \S \ref{sec:summary}.

Throughout this paper, we use the time of nova outburst, 2021 Aug 08.5 UT \citep{munari2021} as the reference time. All times $t$ mentioned are with respect to this reference epoch. 

\section{UGMRT Observations and Data Reduction}
\label{sec:obs}
uGMRT observations of RS Ophiuchi were carried out from 2021 August 31 ($t \sim$ 23 days) to 2022 May 23 ($t \sim$ 287 days) under Director’s Discretionary Time (DDTC204, DDTC205, DDTC207, DDTC211, and DDTC222) and regular GTAC proposals (42$_{-}$083). The observations were done in band-2 (125$-$250 MHz), band-3 (250$-$500 MHz), band-4 (550$-$850 MHz), and band-5 (1000$-$1460 MHz). The data were recorded with an integration time of 10 seconds in the standard continuum mode. We used a processing bandwidth of 400 MHz in bands 4 and 5, 200 MHz in band 3, and 100 MHz in band 2, split into 2048 channels. 3C286, 3C48, and 3C147 were used as flux density calibrators. J1822$-$096 and J1743$-$038 were used as the phase calibrators. 

The Astronomical Image Processing Software \citep[{\sc{aips}};][]{greisen2003} was used for data reduction, following standard radio continuum imaging analysis procedures \citep{Nayana2017}. The data were initially inspected for non-working antennae and radio frequency interference (RFI) affected channels and corrupted data were flagged. The calibration was done using a central channel and the solutions were applied to the entire band. The fully calibrated source data were imaged using {\sc aips} task IMAGR. We performed a few rounds of phase-only self-calibration to improve the image quality. 

We detected radio emission from RS Ophiuchi at all frequencies and the emission was unresolved in all the maps. We estimated the flux density and error of the target source by fitting a Gaussian at the source position using {\sc aips} task JMFIT. The size of the fitted Gaussian is consistent with that of a point source in all the maps. Resolutions of the radio maps are typically 5 $\times$ 2, 7 $\times$ 5, 13 $\times$ 7, and 25 $\times$ 17 arcsec$^{2}$ in bands-5, 4, 3, and 2, respectively. In addition to the errors on flux densities from task JMFIT, we add (in quadrature) a 10\% error in bands-3, 4, and 5 and a 15\% error in band-2 to account for the calibration uncertainties. The flux densities at multiple epochs are listed in Table \ref{tab:gmrt}.

\section{Radio light curves and spectra}
\label{sec:radio-lcs-spectra}
\begin{figure*}
 \includegraphics[scale=0.53]{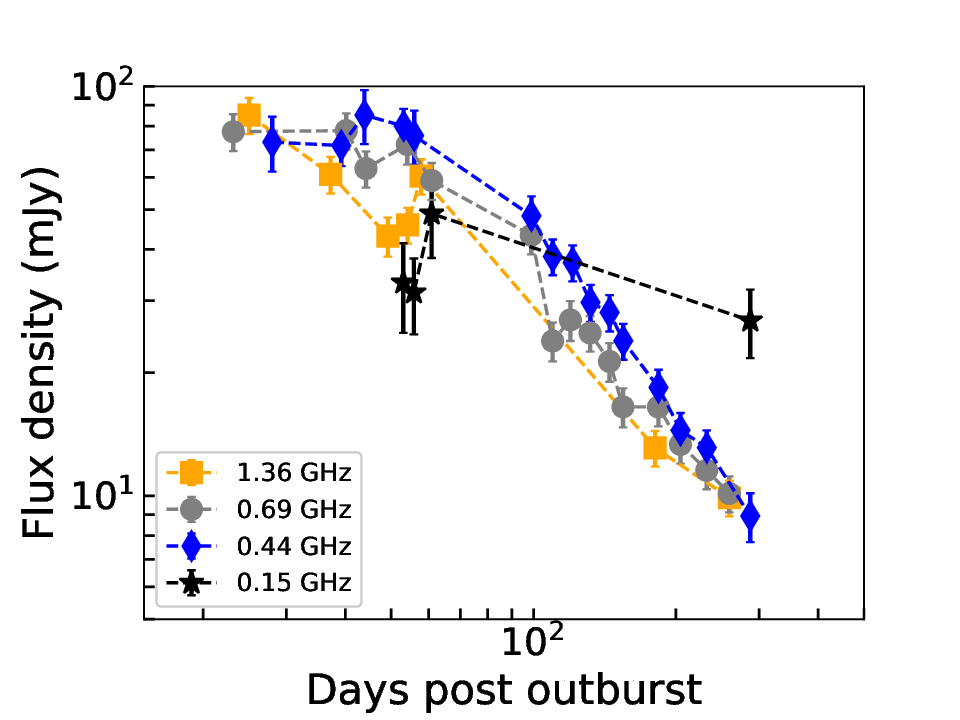} 
\includegraphics[scale=0.50]{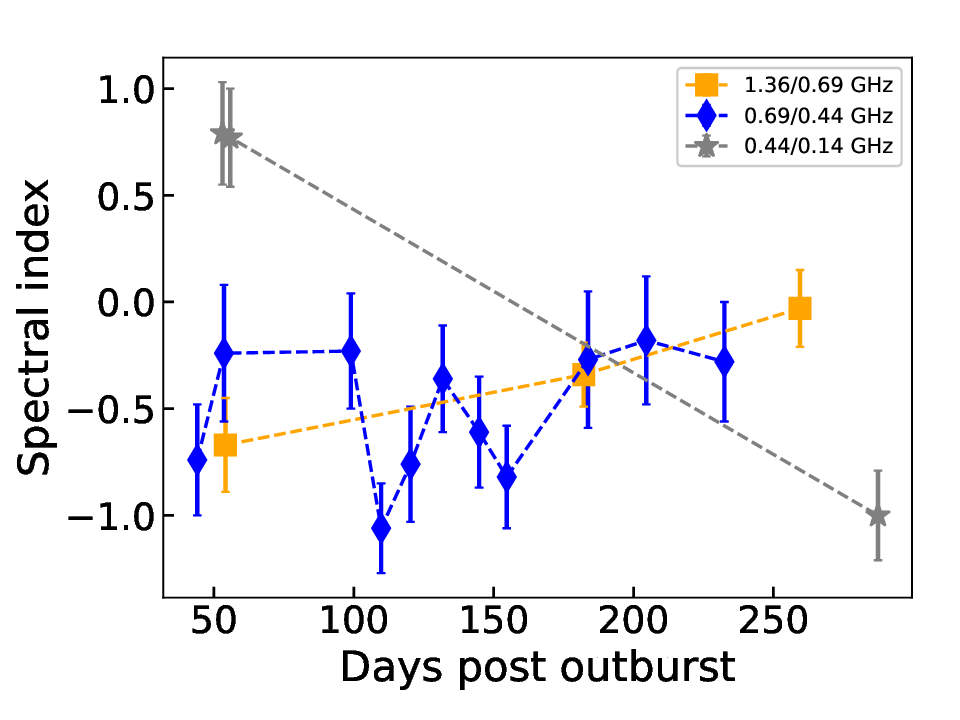} 
    \caption{Left panel: The uGMRT light curves of RS Ophiuchi in its 2021 outburst at 1.36, 0.69, 0.44, and 0.15 GHz. Right panel: The near simultaneous spectral indices between 1.36/0.69 GHz, 0.69/0.44 GHz, and 0.44/0.15 GHz.}
    \label{fig:lc-gmrt-si}
\end{figure*}

The uGMRT light curves at 1.36, 0.69, 0.44, and 0.15 GHz are presented in Fig.~\ref{fig:lc-gmrt-si} (left panel). The 1.36 and 0.69 GHz light curves show a steady decline from a possible maximum at $t \lesssim$ 20 days. The 0.44 GHz light curve peaks slightly later (at $t \sim$ 35 days) than that of the 1.36 and 0.69 GHz light curves. The first three epochs of 0.15 GHz flux measurements are possibly indicating a rising phase, whereas the light curve is in the declining phase at $t \sim 287$ days. However, we note that the error bars on these flux measurements are large. \cite{deruiter2023} presented LOFAR (54 and 154 MHz) and MeerKAT (0.82 and 1.28 GHz) observations of RS Ophiuchi during $t \sim$ 2$-$223 days post the outburst. The combined GMRT+LOFAR+MeerKAT observations suggests the radio light curve evolution to be an initial steep rise to peak followed by a steady decline where higher frequencies trace the peak first (see Fig.~\ref{fig:fit}). This kind of light curve evolution is similar to that seen in radio supernovae \citep{weiler2002}.  

The 1.28 GHz light curve presented in \cite{deruiter2023} shows a plateau between $t \sim 10$ to 40 days. Our band-5 light curve does not show such a flattening towards the peak due to sparse temporal sampling, covering only two measurements between 10 and 40 days.

The near-simultaneous spectral indices, $\alpha$ ($F \propto \nu^{\alpha}$), are estimated at multiple epochs (see Fig.~\ref{fig:lc-gmrt-si} and Table~ \ref{tab:spectral indices}). The spectral index between 1.36/0.69 GHz is $-$0.67 $\pm$ 0.22 at $t \sim$ 54 days and gradually approaches a value $\alpha$ $\sim$ $-$ 0.03 by $t \sim$ 259 days. The average value of spectral index between 0.69 and 0.44 GHz is $\sim -0.6$ during $t \sim 44-183$ days, and flattens to $\alpha \sim -0.2$ to $-0.3$ during $t \sim 204-232$ days. The 0.44/0.15 GHz spectral indices are $\sim$ 0.8 in the rising phase of the light curve and then become $\alpha \sim- 1$ at $t \sim  287$ days. The negative spectral indices suggest that the radio emission is predominantly non-thermal (synchrotron) at the observed frequencies. However, the contribution from non-thermal processes seems to decrease at epochs $t \gtrsim 204$ days. This could possibly be due to the shock deceleration and the shock expanding to a rarified CBM further away from the binary system.

\section{Brightness Temperature}
\label{sec:Tb}
Synchrotron radio emission is characterized by high brightness temperature ($T_{\rm B}$ $\gtrsim$ 10$^{5}$ $-$ 10$^{6}$ K), significantly above the maximum brightness temperature of a photo ionized thermal gas (10$^{4}$ K). We calculate the brightness temperature of the observed radio flux densities using the following equation \citep{chomiuk2021}
\begin{eqnarray}
   \frac{T_{\rm B}}{(K)} = 1200 \left( \frac{F_{\nu}}{\rm mJy} \right) \left( \frac{\nu}{\rm GHz} \right)^{-2} \left( \frac{\theta}{\rm arcsec} \right)^{-2}
\end{eqnarray}

Here, $F_{\nu}$ is the radio flux density at frequency $\nu$, and $\theta$ is the angular size of the radio-emitting region. Estimation of the brightness temperature requires knowing the velocity of the ejected material and hence the geometry and kinematics of the ejecta. High-resolution HST and radio imaging of both the 2006 and 2021 outbursts show that the nova ejecta is bipolar (or equivalently hourglass) in shape \citep{munari2022vlbi,obrien2006,bode2007}. Specifically, radio imaging of the 2021 eruption \citep{munari2022vlbi} supports a geometric model wherein the leading edges of the bipolar lobes have a space velocity of 7550 km/s at $t \sim$ 34 days. In a bipolar morphology, since the velocity flow is expected to be homologous (i.e., $v \propto r$), the leading edge of the bipolar lobes (having  the largest r) has the highest velocity, while the other material, especially the constricted material at the waist, has lower velocities. As a rough approximation, we assume the waist is constricted to the extent that the material there has a space velocity of $\sim$ 3780 km/s i.e. half of that of the leading edge which has a velocity of 7550 km/s. We thus adopt a simplified, mean space velocity of 5660 km/s for the entire ejecta at $t \sim$ 34 days. \cite{mondal2018} reported a velocity deceleration index of the nova ejecta $v \propto t^{-0.66}$ at $t \sim$ 5 to 70 days during the 2006 outburst of RS Ophiuchi. The velocity remains roughly constant up to $t \sim 247$ days after a slight increase around $t \sim$ 80 days.  We use $\theta$ from VLBI observations \citep{munari2022vlbi} at $t \sim$ 34 days and use the velocity deceleration indices from \cite{mondal2018} to calculate the angular size at later epochs. The brightness temperatures corresponding to our flux density measurements are $\sim 10^{5}-10^{8}$ K for an adopted distance of 2.68$^{+0.17}_{-0.15}$ Kpc \citep[Gaia DR3;][]{bailer2021}.
Based on the estimated spectral indices and high brightness temperatures, we conclude that the radio emission is predominantly non-thermal at the observed frequencies.

\begin{table}
	\centering
	\caption{Details of the uGMRT observations of RS Ophiuchi in its 2021 outburst.}
	\label{tab:gmrt}
	\begin{tabular}{lcccr} 
		\hline
		Date of observation & Time$^{a}$ & Frequency & Flux density  \\
(UT) & (Days) & (GHz) & (mJy)\\
		\hline
		2021 Sep 02.53   & 25.03 & 1.36 & 85.10 $\pm$ 8.60   \\ 
 2021 Sep 14.73  & 37.23 & 1.37 & 61.00 $\pm$ 6.30   \\
 2021 Sep 26.69  & 49.19 & 1.36 & 43.10 $\pm$ 4.70   \\
 2021 Sep 31.67  & 54.17 & 1.36 & 45.90 $\pm$ 4.60   \\
 2021 Oct 05.34  & 57.84 & 1.36 & 60.40 $\pm$ 6.00   \\
 2022 Feb 05.26  & 180.76 & 1.36 & 13.10 $\pm$ 1.31  \\
 2022 Apr 25.03  & 259.53 & 1.36 & 09.90 $\pm$ 0.99   \\
 \hline
 2021 Aug 31.67  & 23.17  &  0.69   & 77.50 $\pm$ 8.00  \\
 2021 Sep 17.67  & 40.17  &  0.69   & 78.00 $\pm$ 7.90  \\
 2021 Sep 21.72  & 44.22  &  0.69   & 63.00 $\pm$ 6.40  \\
 2021 Oct 01.58  & 54.08  & 0.69    & 72.10 $\pm$ 7.70  \\
 2021 Oct 08.41  & 60.91  & 0.66    & 58.90 $\pm$ 6.10  \\
 2021 Nov 15.42  & 98.92  & 0.64    & 43.30 $\pm$ 4.40  \\
 2021 Nov 26.25  & 109.75 & 0.69    & 23.90 $\pm$ 2.60  \\
 2021 Dec 06.15  & 119.65 & 0.69    & 26.90 $\pm$ 3.00   \\
 2021 Dec 18.21  & 131.71 & 0.69    & 25.00 $\pm$ 2.50   \\
 2021 Dec 31.26  & 144.76 & 0.69    & 21.30 $\pm$ 2.30  \\
 2022 Jan 10.08  & 154.58 & 0.69    & 16.50 $\pm$ 1.80  \\
 2022 Feb 08.19  & 183.69 & 0.69 & 16.49 $\pm$ 1.71   \\
 2022 Mar 01.04  & 204.54 & 0.69 & 13.36 $\pm$ 1.37   \\
 2022 Mar 28.95  & 232.45 & 0.69 & 11.55 $\pm$ 1.17   \\
 2022 Apr 24.98  & 259.48 & 0.62  & 10.14 $\pm$ 1.02  \\
 \hline
 2021 Sep 05.51 & 28.01   & 0.40    & 73.10 $\pm$ 11.20   \\
 2021 Sep 16.70 & 39.20   & 0.46     &  71.70 $\pm$ 7.80    \\
 2021 Sep 21.38 & 43.88   & 0.46    &  85.10 $\pm$ 12.80   \\
 2021 Sep 30.62 & 53.12   & 0.45    &  80.00 $\pm$ 8.10    \\
 2021 Oct 03.42 &  55.92  & 0.46    & 75.80 $\pm$ 11.50   \\
 2021 Nov 15.49 &  98.99  & 0.40    & 48.30 $\pm$ 05.60   \\
 2021 Nov 26.33 &  109.83 & 0.44    & 38.43 $\pm$ 3.85   \\
 2021 Dec 07.46 &  120.96 & 0.45    & 37.15 $\pm$ 3.74    \\
 2021 Dec 18.46 & 131.96  & 0.43    & 29.68 $\pm$ 3.07    \\
 2021 Dec 31.34 & 144.84  & 0.44    & 28.06 $\pm$ 2.86   \\
 2022 Jan 10.32 & 154.82  & 0.44    & 23.91 $\pm$ 2.41  \\
 2022 Feb 08.00 & 183.81   & 0.46    & 18.40 $\pm$ 1.93   \\
 2022 Mar 01.12 & 204.62   & 0.44    & 14.46 $\pm$ 1.48   \\
 2022 Mar 29.04 & 232.54   & 0.44    & 13.12 $\pm$ 1.34  \\
 2022 May 22.85 & 287.35   & 0.44    & 08.93 $\pm$ 1.22  \\
 \hline
  2021 Sep 30.53  & 53.03  &  0.147    &  33.20 $\pm$ 8.20     \\
 2021 Oct 03.34  & 55.84  &  0.147    &  31.40 $\pm$ 6.60     \\
 2021 Oct 08.38  & 60.88  & 0.146     &  48.90 $\pm$ 10.80     \\
 2022 May 22.91  & 287.41 & 0.147     & 26.80 $\pm$ 5.10     \\
		\hline
	\end{tabular}
 
 \scriptsize{a This column indicates the time from the nova outburst i.e., 2021 Aug 08.5 UT \citep{munari2021}.}
\end{table}

\begin{table*}
	\centering
	\caption{Near simultaneous radio spectral indices of RS Ophiuchi in its 2021 outburst.}
	\label{tab:spectral indices}
	\begin{tabular}{ccccc} 
		\hline
		Time$^{a}$ & & Spectral Indices  \\
  \cline{2-4}
(Day) & (1.4 and 0.69 GHz) & (0.69 and 0.44 GHz) & (0.44 and 0.15 GHz)\\
		\hline
54.13  & $-$0.67  $\pm$ 0.22 & - & - \\
182.22 & $-$0.34  $\pm$ 0.15 & - & - \\
259.51 & $-$0.03  $\pm$ 0.18 & - & - \\
\hline
44.05  & - &  $-$0.74  $\pm$  0.26 & - \\
53.60  & - &  $-$0.24  $\pm$  0.32 & - \\
98.96  & - &  $-$0.23  $\pm$  0.27 & - \\
109.79 & - &  $-$1.06 $\pm$  0.21 & - \\
120.31 & - &  $-$0.76 $\pm$  0.27 & - \\
131.84 & - &  $-$0.36 $\pm$  0.25 & - \\
144.80 & - &  $-$0.61 $\pm$  0.26 & - \\
154.70 & - &  $-$0.82 $\pm$  0.24 & - \\
183.75 & - &  $-$0.27 $\pm$  0.32 & - \\
204.58 & - &  $-$0.18 $\pm$  0.30 & - \\
232.49 & - &  $-$0.28 $\pm$  0.28 & - \\
\hline
53.07  & - & - & 0.79 $\pm$ 0.31 \\
55.88  & - & - & 0.77 $\pm$ 0.45 \\ 
287.38 & - & - & $-$1.00 $\pm$ 0.26  \\
		\hline
	\end{tabular}
 
 \scriptsize{a This column indicates the time from the nova outburst i.e., 2021 Aug 08.5 UT \citep{munari2021}.}
\end{table*}

\section{Radio emission model}
\label{sec:radio-model}
Low-frequency radio light curves of RS Ophiuchi can be represented by adopting the synchrotron emission model from \cite{weiler2002}. According to this model, the outburst ejecta plows into the dense red giant wind driving a forward shock. Particles are accelerated at the shock via diffusive shock acceleration and emit synchrotron radiation. Initially, the emission is suppressed via free-free absorption by a homogeneous and clumpy circumbinary medium in the line of sight. 
The spectral and temporal evolution of radio flux densities can be represented as
\begin{eqnarray}
    F \rm (\nu,t) = K_{\rm 1} \left( \frac{\nu}{1\, \rm GHz} \right)^{\alpha} \left( \frac{t}{20\,\rm days}\right)^{\beta} \nonumber \\ \times \rm exp (-\tau_{\rm homog}) \left[ \frac{1-exp(-\tau_{\rm clumps})}{\tau_{\rm clumps}}\right]
    \label{eqn:flux}
\end{eqnarray}
Here, $K_{\rm 1}$ denotes the flux density normalization constant and $t$ denotes the time since the outburst in days. $\beta$ is the temporal decay index of flux densities in the optically thin regime. $\tau_{\rm homog}$ and $\tau_{\rm clumps}$ represent the optical depths due to homogeneous and clumpy circumbinary medium, respectively.

\begin{eqnarray}
    \tau_{\rm homog} = K_{\rm 2} \left( \frac{\nu}{1 \, \rm GHz} \right)^{-2.1} \left( \frac{t}{20\, \rm days} \right)^{\delta}
    \label{eqn:tau-homo}
\end{eqnarray}

\begin{eqnarray}
    \tau_{\rm clumps} = K_{\rm 3} \left( \frac{\nu}{1 \, \rm GHz} \right)^{-2.1} \left( \frac{t}{20\, \rm days} \right)^{\delta^{'}}
    \label{eqn:tau-clumpy}
\end{eqnarray}
In the above equations, $K_{2}$ and $K_{3}$ are the optical depths due to homogeneous and clumpy circumbinary medium, respectively at 1 GHz on $t =$ 20 days. $\delta$ and $\delta^{'}$ denote the temporal indices of optical depths due to homogeneous and clumpy absorbing medium, respectively.

\begin{figure*}
\includegraphics[scale=0.4]{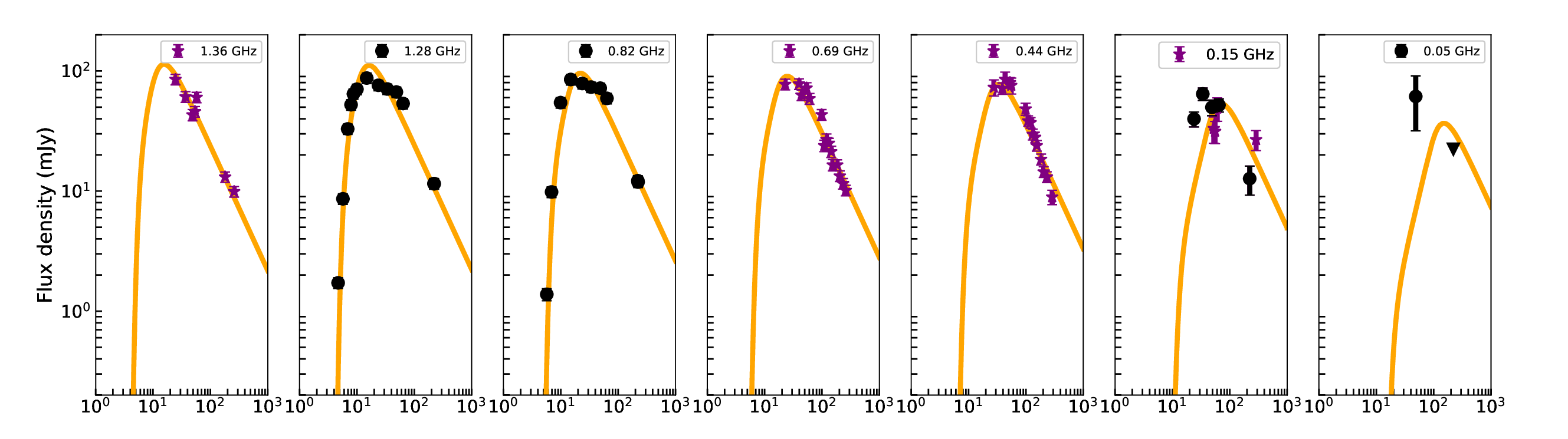} 
\includegraphics[scale=0.4]{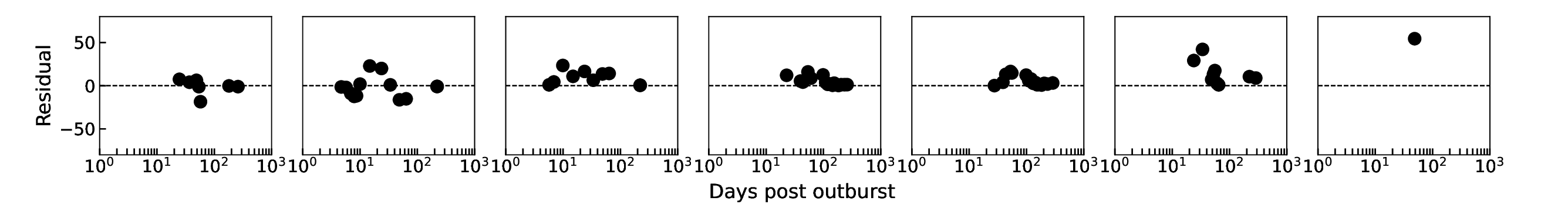} 
\caption{Low-frequency radio light curves of RS Ophiuchi in its 2021 outburst. The data points include flux density measurements from our uGMRT observations and the measurements reported by \protect\cite{deruiter2023} at $t >$ 5 days. The purple star symbols denote the uGMRT flux density measurements and the black circles denote the flux measurements reported by \citet{deruiter2023}. The orange solid curves represent the best-ﬁt single component synchrotron emission model \citep{weiler2002} as described by equations \ref{eqn:flux}, \ref{eqn:tau-homo}, and \ref{eqn:tau-clumpy} (see \S \ref{sec:radio-model}). The bottom panels show the residuals defined as $F_{o}-F_{\rm mod}$, where $F_{\rm o}$ is the observed flux density and $F_{\rm mod}$ is the modeled flux value.} 
\label{fig:fit}
\end{figure*}

We model the radio flux density measurements using equations~(\ref{eqn:flux}, \ref{eqn:tau-homo}, and \ref{eqn:tau-clumpy}) keeping $K_{1}$, $K_{2}$, $K_{3}$, $\alpha$, $\beta$, $\delta$, and $\delta^{'}$ as free parameters. In addition to the uGMRT flux measurements, data from \cite{deruiter2023} are also used. Since it is likely that data at $t <$ 5 days are due to a residual synchrotron component from previous outbursts \citep{deruiter2023}, only the data from $t >$ 5 days are included for modeling. The python \textit{emcee} package \citep{foreman2013} that adopts Markov chain Monte Carlo (MCMC) method is used to execute the fit. We chose 5000 steps and 32 walkers to explore the parameter space to get the best-fit parameters (68\% confidence interval). The best-fit model along with the observed flux values are plotted in Fig.~\ref{fig:fit}  and the best-fit parameters are listed in Table~\ref{tab:model-para}. The reduced chi-square value for the best-fit model is $\chi_{\mu}^{2}=3.7$. We also modeled the data including the flux measurements at t < 5 days and the fit resulted in a much higher reduced $\chi_{\mu}^{2}=6.4$.

The best-fit optically thin spectral and temporal indices are $\alpha = -0.37^{+0.05}_{-0.04}$ and $\beta =-1.04^{+0.03}_{-0.03}$, respectively, roughly similar to the values seen in radio supernovae \citep{weiler2002}.  
The best-fit optical depth normalization parameters due to uniform and clumpy medium are $K_{\rm 2}=1.68 \times 10^{-3}$ and $K_{\rm 3}=0.68$, respectively. These values suggest that a clumpy medium in the line of sight is dominating the absorption processes. We repeated the modeling using only GMRT flux measurements and the best-fit parameters are $K_1 = 171^{+11}_{-9}$ $K_2 = 0.04^{+0.02}_{-0.02}$, $K_3 = 1.33^{+0.35}_{-0.25}$, 
$\alpha = -0.50^{+0.06}_{-0.06}$, 
$\beta = -1.16^{+0.04}_{-0.04}$,
$\delta = -1.74^{+0.52}_{-0.53}$, and 
$\delta^{'} = -3.48^{+0.43}_{-0.38}$.
The values of $K_{1}$, $\alpha$, $\beta$, and $\delta^{'}$ are roughly similar if we use either GMRT or GMRT+LOFAR+MeerKAT data. However, the value of optical depth normalization parameters ($K_{2}$, $K_{3}$) and $\delta$ are significantly different. This indicates that including early time data is critically important to pin down the dominant absorption processes.

While the synchrotron model broadly represents the spectral and temporal behavior of radio light curves, the plateauing towards the peak is not well fitted in the model. 

\begin{figure*}
\begin{centering}
\includegraphics[scale=0.38]{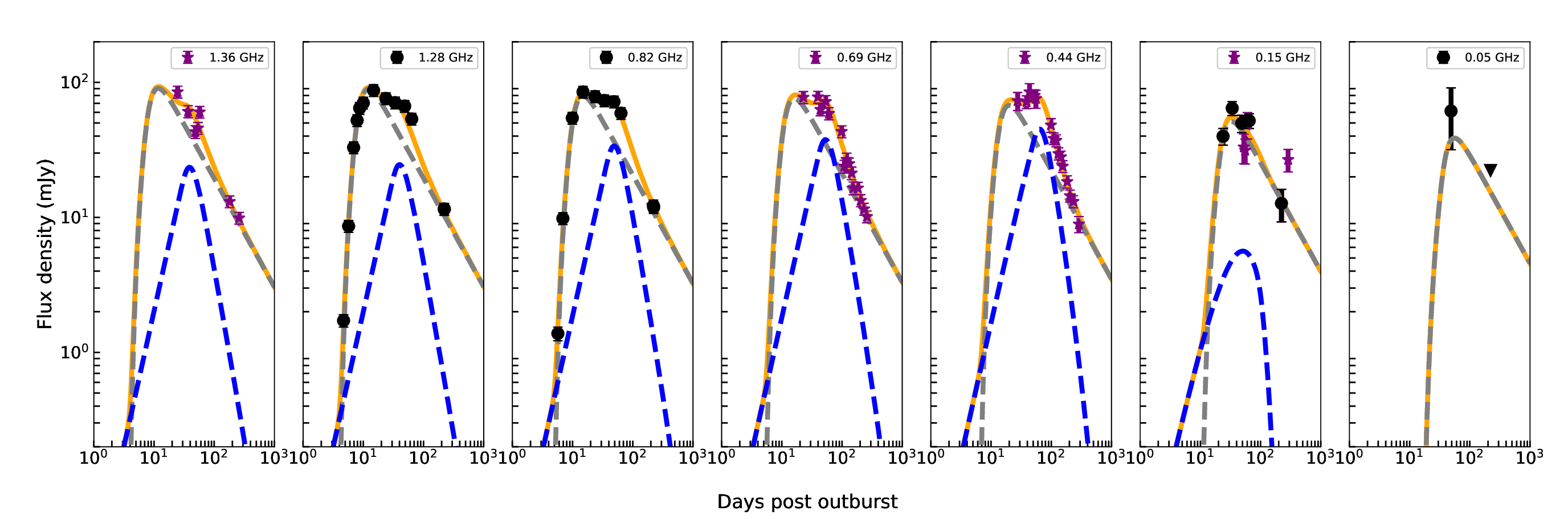} 
\caption{Low-frequency radio light curves of RS Ophiuchi in its 2021 outburst. The data points include flux density measurements from our uGMRT observations and the measurements reported by \protect\cite{deruiter2023} at $t >$ 5 days. The purple star symbols denote the uGMRT flux density measurements and the black circles denote the flux measurements reported by \citet{deruiter2023}. The grey and blue dashed curves represent the first and second components of the synchrotron emission model \citep{weiler2002}, respectively. The orange solid curves represent the total fit which is the sum of two components. The second component of the 0.05 GHz light curve is out of the plot range (the modeled flux values are too low; $F \ll$ 1 mJy). } 
\label{fig:two-compo}
\end{centering}
\end{figure*}

We thus modeled the data in a two-component scenario and found that the fit improved with a reduced chi-square $\chi_\mu^{2} =1.5$. The best-fit model along with the observed flux density measurements are shown in Fig.~\ref{fig:two-compo}. The best-fit parameters that represent each of the synchrotron components are; $K_{1} \sim 60.3$, $K_{2}\sim 1.2\times 10^{-4}$, $K_{3} \sim 2.0 \times 10^{-3}$, $\alpha \sim -0.12$, $\beta \sim -0.8$, $\delta \sim -4.8$, and $\delta^{'} \sim -4.4$ for the first component, and $K_{1} \sim 18.4$, $K_{2} \sim 4.6 \times 10^{-4}$, $K_{3} \sim 1.8, \alpha \sim -1.9$, $\beta \sim -2.5$, $\delta \sim -1.1$, and $\delta^{'} \sim -4.5$ for the second component. The flattening or marginal flux density enhancement close to the peak of the light curves is well represented in the two-component model. The first synchrotron component peaks at $t \sim 12-20$ days, whereas the contribution from the second component is maximum at $t \sim 24-45$ days. The flux density enhancement in the light curves above the prediction of a single (first component) synchrotron model during $t \sim 24-45$ days could be due to a density enhancement in the CBM, as also seen in the 2006 outburst \citep{kantharia2007}. Alternatively, it could also be a consequence of the complex velocity changes of the ejecta as the shock decelerates, as seen for example during the first 50 days of the 2006 eruption \citep[Fig.2;][]{das2006}. Velocity changes would result in fluctuations in the kinetic temperature of the shocked gas which would, in turn, affect the non-thermal emission.

An alternative scenario could involve the emergence of additional radio outflow components to the visibility during $t \sim$ 24$-$45 days. The EVN observations of RS Oph at $t \sim$ 34.3 days \citep{munari2022vlbi} showed three components; one compact central component (CC) and two lobes towards the east and west of the CC \citep[see Fig 2 of][]{munari2022vlbi}. According to the model proposed by \citet{munari2022vlbi}, the eastern arc (EA) remains hidden from the observer at early times due to the high free-free opacity caused by the density enhancement in the orbital plane (DEOP). At times $t > t_{\tau}$ (where $t_{\tau}$ is the time required for EA to expand from the central binary such that the optical depth due to DEOP is $\sim$ 1), the radio emission from EA becomes detectable. As more and more of EA component becomes visible to the observer, the detected flux density will increase. In the 2006 outburst, it was believed that EA became visible during  $t \sim$ 21--51 days \citep{obrien2006,obrien2008,rupen2008,sokoloski2008}. In the 2021 outburst, the EA is in the emerging phase (outer arc is visible) at t $\sim$ 34.3 days \citep{munari2022vlbi}. Considering the similarities between the 2006 and 2021 outbursts in terms of shell velocity and radio morphology, it is very likely that the enhanced flux densities during $t \sim$ 24$-$45 days are due to the eastern lobe becoming visible to the observer.

\cite{kantharia2007} presented low-frequency radio light curves of the 2006 outburst of RS Ophiuchi. We use these flux density measurements and model them (in a single component) following the same method as discussed here. The best-fit values are given in Table~\ref{tab:model-para}. We use the results from the single-component modeling and attempt a detailed comparison of the best-fit parameters with those of the 2006 outburst in \S \ref{sec:comparison-1985,2006 and 2021}.

\begin{table}
	\centering
	\caption{Best fit parameters of the synchrotron emission model.}
	\label{tab:model-para}
	\begin{tabular}{cc} 
		\hline
		2021 outburst  & 2006 outburst \\
		\hline
		$K_1 = 141.25^{+6.97}_{-7.17}$     &
$K_1 = 79.77^{+14.27}_{-18.55}$ \\
$K_2 = 1.68^{+0.78}_{-0.55} \times$ 10$^{-3}$    & 
$K_2 = 0.07^{+0.07}_{-0.06} $ \\
$K_3 =  0.68^{+0.10}_{-0.10}$ &
$K_3 = 0.53^{+0.37}_{-0.29}$    \\
$\alpha = -0.37^{+0.05}_{-0.04}$  & 
$\alpha = -0.73^{+0.14}_{-0.13}$ \\
$\beta = -1.04^{+0.03}_{-0.03}$   & 
$\beta = -1.19^{+0.14}_{-0.09}$ \\
$\delta = -5.74^{+0.28}_{-0.30}$  & 
$\delta = -2.32^{+0.66}_{-0.61}$ \\
$\delta^{'} = -3.05^{+0.10}_{-0.12}$  & 
$\delta^{'} = -3.51^{+0.58}_{-0.65} $ \\
\hline
$\chi^2 = 3.70$                  & 
$\chi^2 = 2.72 $ \\
		\hline
	\end{tabular}

 \scriptsize{$K_{1}$, $K_{2}$, $\alpha$, $\beta$, $\delta$, and $\delta^{'}$ are best-fit parameters of radio emission model defined by equations \ref{eqn:flux}, \ref{eqn:tau-homo} and \ref{eqn:tau-clumpy} (see \S \ref{sec:radio-model}). The modeling (of 2021 outburst) was done including flux density measurements from our uGMRT observations and the ones reported by \protect\cite{deruiter2023} at $t >$ 5 days.}
\end{table}

\subsection{Mass-loss rate}
\label{sec:mass-loss rate}
Mass-loss rate ($\dot{M}$) of the red giant companion can be estimated from the modeled radio light curves \citep{weiler1986,weiler2002}.
\begin{eqnarray}
    \frac{\dot{M}}{v_{w1}} = 3.0 \times 10^{-6} \left< \tau_{\rm eff}^{0.5} \right> m^{-1.5} \left( \frac{v_{\rm i}}{10^{4} \rm km\,s^{-1}} \right)^{1.5} \nonumber
    \\ \times \left( \frac{t_{\rm i}}{45 \rm days} \right)^{1.5} \left( \frac{t}{t_{\rm i}} \right)^{1.5m} \left( \frac{T}{10^{4}\rm K} \right)^{0.68} 
    \label{eqn:mass-loss rate}
\end{eqnarray}

Here, $v_{w1}$ is the stellar wind velocity of the red giant companion in units of 10 km\,s$^{-1}$. $v_{\rm i}$ is the ejecta velocity at $t_{\rm i}$ days post-outburst. $m$ and $T$ represent the shock deceleration parameter ($m=\delta/3$) and wind electron temperature, respectively. $\tau_{\rm eff}$ is the effective optical depth due to a uniform and clumpy absorbing medium \citep{weiler2002}.

\begin{eqnarray}
   \left< \tau_{\rm eff}^{0.5} \right> = 0.67 \left[ (\tau_{\rm homog} + \tau_{\rm clumps})^{1.5} - \tau_{\rm homog}^{1.5} \right] \tau_{\rm clumps}^{-1}
    \label{eqn:tau-eff}
\end{eqnarray}

We estimate the mass-loss rate to be $\dot{M} =$ 7.5$^{+1.5}_{-1.6}$ $\times$ 10$^{-8}$ $M_{\odot}$\,yr$^{-1}$ for a wind velocity of 20 km\,s$^{-1}$ and $T =$ 20,000 K \citep{weiler2002}. We use $v_{\rm i} \sim$ 5660 km\,s$^{-1}$ at $t_{\rm i} =$ 34 days (see \S \ref{sec:Tb}). The derived mass-loss rate is in agreement with the typical mass-loss rates (10$^{-9}$ $-$ 10$^{-6}$ $M_{\odot}$\,yr$^{-1}$) of red giant branch (RGB) stars \citep{vanloon2006,origlia2007,mcdonald2007} and with that presented in \citep{deruiter2023}. The above-mentioned formulation of mass-loss rate is originally developed for supernovae \citep{weiler2002} assuming a spherical geometry. In the case of RS Oph, the ejecta is bipolar and the estimated $\dot{M}$ value could be a rough approximation.

\section{Comparison between the radio evolution of 1985, 2006, and 2021 outbursts of RS Ophiuchi}
\label{sec:comparison-1985,2006 and 2021}
Radio emission was detected and monitored from the 1985 \citep{hjellming1986} and 2006 \citep{kantharia2007,eyres2009} outbursts of RS Ophiuchi. In this section, we compare the properties of radio emission from previous outbursts with that of 2021 to understand the evolution of the system in terms of shock energetics and the circumbinary medium.

\cite{hjellming1986} presented radio observations spanning $t \sim$ 29 $-$ 370 days post the 1985 outburst at frequencies 1.49, 4.85, 14.94, and 22.46 GHz with the Very Large Array (VLA). Radio emission from the 2006 outburst of RS Ophiuchi was detected at frequencies 0.24, 0.33, 0.61, 1.49, 4.89, 6.0, 14.96, and 22.48 GHz \citep{narumi2006,eyres2009,kantharia2007}. In both outbursts, two spectral components were identified, one with a negative spectral index between 1.4/5 GHz and below, and the other with a positive spectral index above 5 GHz. 
The GMRT observations of the 2006 outburst were carried out by \cite{kantharia2007} during $t \sim$ 13 $-$ 110 days post-discovery at frequencies 1.39, 1.28, 1.06, 0.61, 0.33, 0.24, and 0.15 GHz. Radio emission was detected at all frequencies except at $\nu =$ 0.15 GHz at $t \sim$ 110 days with a 3$\sigma$ flux density limit of 27 mJy. 

\begin{figure*}
\begin{centering}
\includegraphics[scale=0.45]{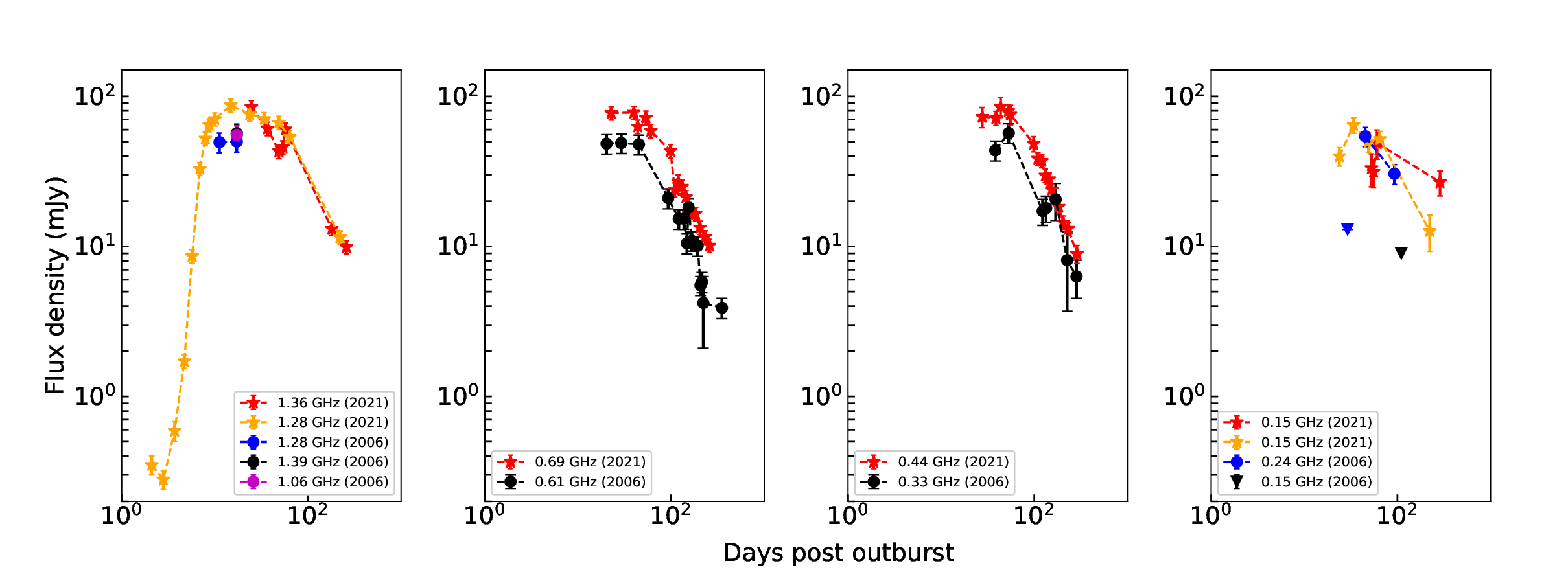} 
\caption{Radio light curves of RS Ophiuchi in its 2006 and 2021 outburst. The uGMRT radio flux density measurements of the 2021 outburst are marked as red star symbols. The flux densities of the 2021 outburst reported in \protect\cite{deruiter2023} are marked as orange star symbols. The blue and black filled circles and downward triangles represent the flux densities and limits, respectively during the 2006 outburst \citep{kantharia2007}.}
\label{fig:lc-compare-2006-2021}
\end{centering}
\end{figure*}

We compile all published low-frequency ($\nu \leq$ 1.4 GHz) radio observations of 1985 and 2006 outbursts to investigate the evolution of shock-driven synchrotron emission over successive outbursts. Fig.~\ref{fig:lc-compare-2006-2021} shows the radio light curves at $\nu \sim$ 0.15 $-$ 1.4 GHz of RS Ophiuchi in its 2006 and 2021 outbursts and Fig.~\ref{fig:lc-compare-all} shows the 1.4 GHz light curves of 1985, 2006, and 2021 outbursts. 

When we compare the low-frequency radio light curves of the 2006 and 2021 outbursts, two characteristics are particularly striking. 
\begin{enumerate}
    \item The radio flux densities during the 2021 outburst are systematically higher than those of the 2006 outburst.
    \item Radio emission at 0.15 GHz is detected from $t \sim$ 24 to 287 days at 9 epochs during the 2021 outburst. Radio observations at 0.15 GHz were carried out only at a single epoch ($t \sim$ 110 days) in the 2006 outburst which resulted in non-detection with a 3-sigma flux density limit of $F <$ 27 mJy.
\end{enumerate}

\subsection{Brighter synchrotron emission during the 2021 outburst}
\label{subsec:reason-brighter-radio-emission-2021}
The observed peak flux densities of the bands 5, 4, and 3 light curves in the 2021 outburst are $\sim$ 85, 78, and 85 mJy, respectively. The corresponding values in the 2006 outburst are $\sim$ 57, 49, and 57 mJy, respectively \citep{kantharia2007}, about 1.5--1.6 times lower. If we compare the flux densities at higher frequencies between 2021 \citep{sokolovsky2021} and 2006 outbursts \citep{eyres2009}, this trend seems to be broadband.

The key dependencies on the synchrotron emissivity [$J(\nu)$] are \citep[equation 8.89; ][]{longairbook2011}
\begin{equation}
J(\nu) \propto N_{\rm 0}\,B^{(p+1)/2}\, \nu^{-(p-1)/2}
\end{equation}
where the power-law distribution of electron energies is represented as $N(E)dE =$ $N_{\rm 0}E^{-p}dE$. Here, $N(E)dE$ is the number density of electrons of energies between $E$ and $E+dE$. Thus a higher synchrotron radio flux density could be due to more synchrotron emitting electrons in the emitting region and/or due to a higher magnetic field. 

Assuming the amplified post-shock magnetic field is in equipartition with the relativistic electrons in the radio-emitting region \citep{chevalier2006,chugai2010}, the magnetic field can be written as \citep{chomiuk2012} 
\begin{equation}
B =\sqrt{8 \pi \epsilon_{\rm B} \mu m_{\rm H} n_{\rm CBM} v^{2}}
\end{equation}
Here, $n_{\rm CBM}$ is the particle number density in the CBM and $v$ is the shock velocity. \cite{munari2022vlbi} reported the size and expansion velocity of the radio-emitting region at $t \sim$ 34 days in the 2021 outburst and mentioned that the size is remarkably similar in the case of 2006 outburst at similar epochs. The hard X-ray emission and its temporal evolution are also reported to be broadly similar between the two outbursts \citep{page2022}. Both these observations are indicative of similar shock velocities during the 2006 and 2021 outbursts. Thus the higher synchrotron brightness in the 2021 outburst could be due to a higher density of the CBM. The synchrotron radio flux density can be represented in terms of electron number density in the synchrotron plasma as $F \propto$ $n_{\rm e}^{(p+5)/4}$ where $p$ is related to the optically thin spectral index as; $\alpha = (p-1)/2$. Thus a 1.5$-$1.6 times increase in synchrotron flux densities during the 2021 outburst compared to that of 2006 could indicate a $\sim$ 30\% increase in $n_{\rm e}$ considering the shock velocity and magnetic field are the same during both outbursts.

We detect 0.15 GHz radio emission at $t \sim 52, 55, 60, 287$ days post the 2021 outburst of RS Ophiuchi. \cite{deruiter2023} reported radio emission at 0.15 GHz at $ t \sim$ 24, 34, 49, 65, and 223 days post outburst. Observations at this frequency at a single epoch ($t \sim$ 110 days) during the 2006 outburst resulted in non-detection with a 3$\sigma$ flux density limit of 27 mJy. However, during the 2021 outburst, the best fit model predicts a flux density of $\sim$ 44 mJy at 110 days, which is significantly higher than the flux density limit during the 2006 outburst. This could either be indicative of an early turn-on of the 0.15 GHz radio emission in the 2021 outburst or be due to the intrisically brighter synchrotron emission in 2021 compared to that of 2006.

\begin{figure}
\begin{centering}
\includegraphics[scale=0.5]{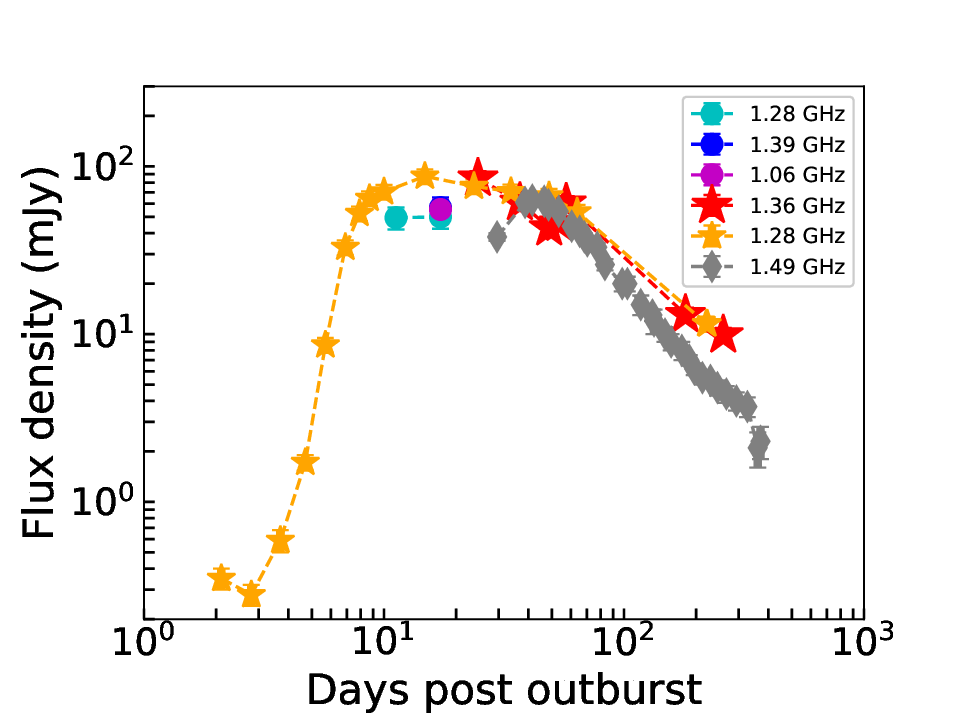} 
\caption{Radio light curves of RS Ophiuchi in its 1985, 2006, and 2021 outburst. The uGMRT radio flux density measurements of the 2021 outburst are marked as red star symbols. The flux densities of the 2021 outburst reported in \protect\cite{deruiter2023} are marked as orange star symbols. The blue, cyan, and magenta filled circles represent the flux densities during the 2006 outburst \citep{kantharia2007}. Grey diamonds denote the flux density measurements from the 1985 outburst \citep{hjellming1986}.}
\label{fig:lc-compare-all}
\end{centering}
\end{figure}

We now look at the published 1.4 GHz radio observations during the 1985 outburst along with those of 2006 and 2021 in Fig.~\ref{fig:lc-compare-all}. The radio emission is brightest in the 2021 outburst whereas the 2006 and 1985 outbursts show roughly similar flux values. The time of peak of the 1.4 GHz light curve ($t_{\rm p}$) is not well constrained in the 1985 outburst due to sparse sampling of the light curve at $t < 20$ days \citep[see Fig 1 of][]{hjellming1986}. We assume the peak of the 1.4 GHz light curve in the 1985 outburst to be $t_{\rm p} \lesssim$ 20 days, similar to that of the 2006 ($t_{\rm p} \sim$ 15 days) and 2021 ($t_{\rm p} \sim$ 15 days) outbursts. The 0.325 GHz emission was not detected from the 1985 outburst at $t \sim$ 48 days down to a flux density limit of 5 mJy whereas 0.325 GHz emission was detected at $t \sim$ 38 days in the 2006 outburst with a flux density of 43.7 mJy. The non-detection of 0.325 GHz emission are indications of higher foreground absorption in 1985 than in 2006 \citep{kantharia2007}.

\subsection{Evolution of optical depths due to uniform and clumpy absorbing medium}
We investigate the evolution of optical depth due to uniform and clumpy components of the absorbing medium during the 2006 and 2021 outbursts from the best-fit modeled parameters. Table~\ref{tab:optical-depths} shows the values of $\tau_{\rm homo}$, $\tau_{\rm clumps}$, and $\tau_{\rm eff}$ in the 2006 and 2021 outbursts at the time of the peak of the 1.36, 0.69, 0.44, and 0.15 GHz light curves in 2021 outburst. The optical depth due to the inhomogeneous component of the absorbing medium is systematically higher in 2021 than in 2006 suggesting that the clumpiness in the circumbinary medium has increased from 2006 to 2021. 

\begin{table*}
	\begin{centering}
	\caption{Optical depth due to uniform and clumpy circumburst medium during the 2021 and 2006 outbursts.}
	\label{tab:optical-depths}
	\begin{tabular}{cccccccc} 
		\hline
		Time$^{a}$ & & 2021 & & & 2006 &  \\
  \cline{2-4}
\cline{4-7}
(Day) & $\tau_{\rm homo}$ & $\tau_{\rm clumps}$ & $\tau_{\rm eff}$ & $\tau_{\rm homo}$ & $\tau_{\rm clumps}$ & $\tau_{\rm eff}$ \\
 & ($\times$10$^{-3}$) & & &  &  & \\
		\hline
15.0 & 1.58 $\pm$ 0.46 & 0.38 $\pm$ 0.06 & 0.17 $\pm$ 0.03 & 0.04 $\pm$ 0.03 & 0.30 $\pm$ 0.21 & 0.18 $\pm$ 0.10  \\ 
24.2 & 0.60 $\pm$ 0.19 & 0.44 $\pm$ 0.07 & 0.20 $\pm$ 0.03 & 0.06 $\pm$ 0.05 & 0.29 $\pm$ 0.21 & 0.20 $\pm$ 0.11 \\ 
33.1 & 0.59 $\pm$ 0.20 & 0.69 $\pm$ 0.11 & 0.31 $\pm$ 0.05 & 0.11 $\pm$ 0.09 & 0.41 $\pm$ 0.33 & 0.30 $\pm$ 0.18  \\ 
70.4 & 0.04 $\pm$ 0.02 & 0.52 $\pm$ 0.12 & 0.23 $\pm$ 0.05 & 0.15 $\pm$ 0.18 & 0.21 $\pm$ 0.24 & 0.25 $\pm$ 0.22  \\ 
		\hline
	\end{tabular}
\end{centering}

\scriptsize{a This column indicates the time from the nova outburst i.e., 2021 Aug 08.5 UT \citep{munari2021}.}

 \scriptsize{$\tau_{\rm homo}$ and $\tau_{\rm clumps}$ denote the optical depths due to uniform and clumpy mediums, respectively. $\tau_{\rm eff}$ is the effective optical depth \citep{weiler2002}.}
\end{table*}

\section{A comprehensive picture} 
\label{sec:comprehensive}
Shock-driven synchrotron radio emission from multiple outbursts of RS Ophiuchi shows differences in radio brightness. It is important to tie down these differences with observational characteristics at other frequencies (optical, X-ray, $\gamma$-rays) and theoretical models to build a comprehensive picture.

The optical light curves and evolution between 2006 and 2021 outbursts are almost identical, indicating that the ejecta mass, the mass of the white dwarf, and ejecta velocities are broadly the same \citep{page2022}. The hard X-ray emission ($>$ 1 keV) during the 2021 and 2006 outbursts are similar in terms of brightness and temperature evolution \citep{orio2022,page2022} suggesting the gross properties of the shock interaction to be similar. The results from VLBI observations also point to the same conclusions. The shock radius at $t \sim$ 34 days in the 2021 outburst from VLBI measurements \citep{munari2022vlbi} is remarkably the same as the shock radius at similar epochs in the 2006 eruption \citep{obrien2006,obrien2008,sokoloski2008,rupen2008}. Contrary to a very similar evolution in optical, hard X-ray, and VLBI observations, the soft ($0.3-1$ keV) X-ray emission during the SSS phase was found to be very different. The SSS emission during the 2021 outburst is $4-5$ times fainter than that of the 2006 outburst \citep{page2022}. 

A higher accretion rate can lead to a rapid increase in temperature of the WD surface for a given accreted mass \citep{prialnik1982} and result in a weaker nova and SSS emission. However, there are no observational signatures of a difference in accretion rate during the quiescent intervals prior to 2006 and 2021 outbursts \citep{orio1993,page2022}. The quiescent X-ray luminosities (0.3 $-$ 10 keV) are $\sim$ 6 $\times$ 10$^{31}$ erg\,s$^{-1}$, orders of magnitude less compared to the expected value of $\sim$ 10$^{36}$ erg\,s$^{-1}$ \citep{osborne2011,anupama1999}. A possible explanation for such suppression in quiescent luminosity is the complex (partially covering) absorbing medium in the system. \cite{ness2023} compared the X-ray grating spectra of 2006 and 2021 outbursts during the SSS phase and concluded that higher absorption in the line of sight is the main reason for the observed lower SSS emission in 2021. 

 The line of sight absorbing material is created due to aspherical mass-loss in the binary system and could have a dependence on the orbital phase \citep{orlando2009,drake2009,walder2008}. The 1985, 2006, and 2021 outbursts occurred at phases 0.32, 0.26, and 0.72, respectively. Simulation results from \cite{booth2016} showed that a spiral accretion during the quiescent phase builds up material such that there could be more material in the line of sight during the 2021 outburst. The simulation shows a low density at an orbital phase $\sim$ 0.26, which is during the 2006 outburst. The authors also suggest that the simulated material is clumpy.  

We re-visit the differences in the evolution of low-frequency radio light curves in 2006 and 2021 outbursts in light of the results from \cite{ness2023}. The radio emission is the brightest in the 2021 outburst with a peak flux density of $\sim$ 87 mJy at 1.4 GHz on $t \sim$ 15 days post outburst \citep{deruiter2023}. The 1.4 GHz light curve is sparsely sampled in the 2006 outburst and the peak flux and time from the best-fit model are $F_{\rm p} \sim$ 61 mJy and $t \sim$ 15 days, respectively \citep{kantharia2007}. Assuming similar ejecta and shock properties, these peak flux densities indicate that there are more particles in the synchrotron emitting plasma in 2021 than that of 2006 outburst. 

The integrated X-ray count from SSS emission in the 2021 outburst during $t \sim$ 30 to 86 days is 4$-$5 times lower than that of 2006 possibly due to higher absorption \citep{ness2023}. These two observational characteristics point out that the medium in the vicinity of the binary system during the 2021 outburst is relatively denser compared to that of 2006 outburst.
\section{Summary}
\label{sec:summary}
We present low-frequency radio observations of the 2021 outburst of RS Ophiuchi during 23 to 287 days post outburst at frequencies 0.15 $-$ 1.4 GHz. Our main findings are the following:
\begin{enumerate}
    \item Radio emission at the observed frequencies are characterized by high brightness temperatures (10$^{5}$ $-$ 10$^{8}$ K) with an average optically thin spectral index of $\alpha \sim$ $-$ 0.4, consistent with non-thermal emission process.
    \item The spectral and temporal evolution of the radio flux measurements are best represented by synchrotron emission that arises due to the interaction of external shock from the nova outburst with the dense wind of the red giant companion. The emission is initially suppressed due to free-free absorption by the uniform and clumpy foreground circumbinary medium.
    \item The mass-loss rate of the red giant has been estimated to be $\dot{M} =$ 7.5 $^{+1.5}_{-1.6}$ $\times$ 10$^{-8}$ $M_{\odot}$\,yr$^{-1}$ for a wind velocity of 20 km\,s$^{-1}$, consistent with the mass-loss rates of RGB stars \citep{mcdonald2007}.
    \item Comparison of light curves from the 2021 and 2006 outbursts in the 0.15$-$1.4 GHz radio bands shows that the peak flux densities in the 2021 outburst are 1.5 $-$ 1.6 times higher than that of the 2006 outburst values. Considering similar shock and ejecta properties of these outbursts as supported by multiwavelength observations \citep{page2022,orio2022,munari2022vlbi}, we interpret the brighter radio emission in 2021 to be due to a higher particle density in the synchrotron emitting plasma. 

 Our study suggests that rapid multi-frequency radio follow-up observations of recurrent novae are crucial to tie down various absorption processes and infer the properties of shock and CBM.    
\end{enumerate}

\section*{Acknowledgements}
We thank the anonymous referee for the comments that helped to improve our manuscript. We thank the staff of the GMRT that made these observations possible. The GMRT is run by the National Centre for Radio Astrophysics of the Tata Institute of Fundamental Research. Nayana A.J. would like to acknowledge DST-INSPIRE Faculty Fellowship (IFA20-PH-259) for supporting this research. KPS thanks the Indian National Science Academy for support under the INSA Senior Scientist Programme.

\section*{Data Availability}
The data underlying this article will be shared on reasonable request to the corresponding author.


\bibliographystyle{mnras}
\bibliography{example} 



\bsp	
\label{lastpage}
\end{document}